\documentclass[aps,prd,twocolumn,showpacs,preprintnumbers,amsmath,amssymb]{revtex4}

\usepackage{graphicx}
\usepackage{dcolumn}
\usepackage{bm}

\begin{document}
\title{A chiral quark model study of $Z^+(4430)$ in the molecular picture}

\author{Yan-Rui Liu $^1$}
\email{yrliu@ihep.ac.cn}

\author{Zong-Ye Zhang $^{1,2}$}
\email{zhangzy@ihep.ac.cn}

\affiliation{1. Institute of High Energy Physics, CAS, P.O. Box
918-4, Beijing 100049, China\\
2. Theoretical Physics Center for Science Facilities, CAS, Beijing
100049, China}

\date{\today}

\begin{abstract}

We investigated the bound state problem of the S wave charged $D_1
\bar{D}^*$ ($D_1' \bar{D}^*$) system in a chiral quark model by
solving the resonating group method equation. Our preliminary study
does not favor the molecular assumption of $Z^+(4430)$. On the
contrary, if $Z^+(4430)$ is really a molecule, its partner with
opposite $G$-parity should also exist and probably may be found in
the $\pi^+\eta_c(2S)$, $J/\psi \pi^+\pi^0$, or $\psi'\pi^+\pi^0$
channel. For the bottom systems, we found the existence of both
$I^G$=$1^+$ and $I^G$=$1^-$ $B_1 \bar{B}^*$ ($B_1' \bar{B}^*$)
molecules is possible.
\end{abstract}


\maketitle

\section{introduction}\label{sec1}

In recent years, a series of heavy quark hadrons with unexpected
properties were observed one by one, from $D_{sJ}(2317)$
\cite{ds2317-babar,ds2317-belle,ds2317-cleo}, $X(3872)$
\cite{3872-Belle,3872-CDF,3872-D0,3872-BaBar}, $Y(4260)$
\cite{4260-BaBar,4260-Cleo,4260-Belle}, $X(3940)$
\cite{x3940-belle}, and $Y(3940)$ \cite{y3940-belle,y3940-babar}, to
the newly observed $Y(4140)$ \cite{4140-CDF}. These near threshold
mesons stimulated the interpretations beyond the quark model. For
the interesting hidden charm XYZ states, the interpretations include
tetraquark or molecular states, hybrid charmonia, cusps, and
threshold effects. However, it is not excluded that these XYZ states
are still mesons dominated by $c\bar{c}$ components. On the
contrary, the observation of charged charmonium-like states is a
surprising issue because such states contain at least four quarks.

The Belle Collaboration announced a distinct peak $Z^+(4430)$ in the
$\pi^+ \psi^\prime$ invariant mass distribution in the decay $B\to
K\pi^\pm \psi^\prime$ in Ref. \cite{4430-belle}. The mass and width
are $M=4433\pm 4(stat)\pm 2(syst)$ MeV and
$\Gamma=45^{+18}_{-13}(stat)^{+30}_{-13}(syst)$ MeV, respectively.
The minimum quark content is $c\bar{c}u\bar{d}$. Very recently, a
little heavier and broader $Z^+(4430)$ is obtained in Belle's
reanalysis based on the same data sample \cite{4430-belle-again}.
However, the experimental data from the BaBar collaboration do not
provide significant evidence for the existence of $Z^+(4430)$
\cite{4430-BaBar}.

In addition to $Z^+(4430)$, Belle Collaboration recently observed
two more charged resonance-like structures in the $\pi^+\chi_{c1}$
invariant mass distribution in $B\to K\pi^\pm \psi^\prime$ decays
\cite{twocharged}. The mass and width for the first structure are
\begin{eqnarray}
&M_1=4051\pm 14(\mathrm{stat})^{+20}_{-41}(\mathrm{syst})\,\mathrm{MeV}&\\
&\Gamma_1=82^{+21}_{-17}(\mathrm{stat})^{+47}_{-22}(\mathrm{syst})\,\mathrm{MeV}&
\end{eqnarray}
while the values for the second one are
\begin{eqnarray}
&M_2=4248^{+44}_{-29}(\mathrm{stat})^{+180}_{-35}(\mathrm{syst})\,\mathrm{MeV}&\\
&\Gamma_2=177^{+54}_{-39}(\mathrm{stat})^{+316}_{-61}(\mathrm{syst})\,\mathrm{MeV}.&
\end{eqnarray}
Whether the existence of these two structures is supported awaits
experimental confirmation from other collaborations.

Since the announcement of $Z^+(4430)$, lots of discussions in
various pictures have appeared, which include a tetraquark state
\cite{4430-maiani,4430-string}, a resonance or a molecule in the
$D_1\bar{D}^*$ ($D_1^\prime \bar{D}^*$) channel
\cite{4430-chao,4430-ding,4430-dinget,4430-lee,4430-xiang}, a
baryonium state \cite{4430-qiao}, a threshold cusp \cite{4430-bugg}
and a radial excited $c\bar{s}$ state \cite{4430-Ds}. In Ref.
\cite{4430-bottom}, the bottom analogs of $Z^+(4430)$ were studied.
Besides the spectroscopy, there were discussions about its
production
\cite{4430-rosner,4430-partner,4430-braaten,4430-production,4430-NNbar}
and decay \cite{4430-width}. The $\pi\psi'$ scattering is studied in
Ref. \cite{4430-scatt}.

The molecular picture is widely used because $Z^+(4430)$ is close to
the threshold of $D_1^\prime D^*$ or $D_1D^*$. In Ref.
\cite{4430-xiang}, the calculation at hadron level indicates that it
is possible to get a bound state in the $D_1^\prime \bar{D}^*$ or
$D_1\bar{D}^*$ system with appropriate parameters and to interpret
$Z^+(4430)$ as a molecule. The QCD sum rule study \cite{4430-lee}
and a quark model calculation \cite{4430-dinget} also favor the
$D_1\bar{D}^*$ molecule interpretation. However, a recent
calculation on the lattice indicates such an interpretation is
probably problematic \cite{lat-4430}.

To help to understand this charged state further, we present our
preliminary study from a chiral quark model ($\chi$QM) \cite{chiQM}
and an extended chiral quark model (E$\chi$QM) \cite{ExchiQM}
calculation in this article. The former model includes $\sigma$ and
$\pi$ exchange interactions between light quarks. The later model is
an extended version of the former one by including the vector meson
exchanges. We investigate the bound state problem of the S-wave
$D_1^\prime \bar{D}^*$ or $D_1\bar{D}^*$ system by solving the
resonating group method (RGM) equation \cite{Oka}. In previous
studies, this approach has successfully reproduced the energies of
the light quark baryon states, the binding energy of the deuteron
and the NN scattering phase shifts. When using it to study the
system of a light meson and a light baryon \cite{NKDeltaK}, the
resulting phase shifts are also in agreement with the experimental
data. With this model, we have preliminarily studied the bound state
problem of two S-wave heavy mesons in Ref. \cite{3872-rgm} and
\cite{hqmole-rgm}. The results are roughly consistent with similar
studies at hadron level \cite{3872-liu,hqmole-xiang}. We here intend
to explore whether or not the model can be used to the case of
orbitally excited mesons.

For the system studied, the orbitally excited heavy mesons are $D_1$
and $D_1^\prime$. These two $J^P$=$1^+$ mesons are mixed from
$^3P_1$ and $^1P_1$ states
\begin{eqnarray}
|D_1 \rangle&=&\cos\theta|^1P_1 \rangle + \sin\theta|^3P_1 \rangle \nonumber\\
|D_1^\prime \rangle&=&-\sin\theta|^1P_1 \rangle + \cos\theta|^3P_1
\rangle.
\end{eqnarray}
The mixing angle $\theta=-54.7^\circ$ or $35.3^\circ$ may be deduced
with the mass of the heavy quark going into infinity. In this
article, we adopt the widely adopted value $\theta=-54.7^\circ$
\cite{4430-dinget,mixangle}.

In the molecular picture, the flavor wave function of $Z^+(4430)$
reads
\begin{equation}
Z^+(4430)=\frac{1}{\sqrt2}(\bar{D}_1^0D^{*+}+\bar{D}^{*0}D_1^+)
\end{equation}
or
\begin{equation}
Z^+(4430)=\frac{1}{\sqrt2}(\bar{D}_1^{'0}D^{*+}+\bar{D}^{*0}D_1^{'+}).
\end{equation}
The quantum numbers are $I^GJ^P$=$1^+(0,1,2)^-$. As a preliminary
study, we consider only interactions involving color-singlet mesons.
The one-meson exchange potentials between the heavy mesons are
induced by the meson exchanges between light quarks.

This paper is organized as follows. After the introduction, we
present our model in Sec. \ref{sec2}. Then in Sec \ref{sec3}, we
present our results and discussions.

\section{The chiral quark model}\label{sec2}

The Hamiltonian for the heavy quark meson-antimeson system in the
chiral quark model has the form \cite{chiQM,ExchiQM}
\begin{eqnarray}\label{ham}
H &=&\sum_{i=1}^4 T_i -T_G + V^{OGE}+V^{conf}+\sum_{M}V^M
\end{eqnarray}
where $T_i$ is the kinetic term of the $i$th quark or antiquark and
$T_G$ is the kinetic energy operator of the center of mass motion.
$M$ is the exchanged meson between light quarks. In the chiral quark
model, one of the sources for the constitute quark mass is the
coupling with chiral fields which come from the spontaneous vacuum
breaking. Because the breaking has small effects on the generation
of the constitute mass of the heavy quarks, the coupling of the
$\sigma$ meson and the heavy quarks should be weak. As a result, the
possible flavor-singlet meson exchange interactions between heavy
quarks and between a heavy quark and a light quark have small
contributions and so we ignore them.

The potential induced by the one-gluon-exchange (OGE) interaction
reads
\begin{eqnarray}
V_{\bar{q}Q}^{OGE}&=&g_qg_Q
\mathbf{F}^c_{\bar{q}}\cdot\mathbf{F}^c_Q\left\{\frac{1}{r}-\frac{\pi}{2}\delta^3(\mathbf{r})\Big[\frac{1}{m_q^2}+\frac{1}{m_Q^2}\right.\nonumber\\
&&\left.+\frac43\frac{1}{m_qm_Q}(\bm{\sigma}_q\cdot\bm{\sigma}_Q)\Big]\right\}+V_{OGE}^{\bf{l}\cdot\bf{s}},\\
V_{OGE}^{\bf{l}\cdot\bf{s}}&=&-\frac14g_qg_Q\mathbf{F}^c_{\bar{q}}\cdot\mathbf{F}^c_Q
\frac{3}{m_qm_Q}\frac{1}{r^3}\bf{L}\cdot(\bm{\sigma}_q+\bm{\sigma}_Q),
\end{eqnarray}
where $\mathbf{F}^c_{Q}=\frac{\bm{\lambda}}{2}$ for quarks and
$\mathbf{F}^c_{\bar{q}}=-\frac{\bm{\lambda}^\ast}{2}$ for antiquarks
and $m_q$ $(m_Q)$ is the light (heavy) quark mass. The linear
confinement potential inside the color-singlet meson is
\begin{eqnarray*}
V_{\bar{q}Q}^{conf}=-4\mathbf{F}^c_{\bar{q}}\cdot\mathbf{F}^c_Q\left(a^c_{qQ}
r +a^{c0}_{qQ}\right).
\end{eqnarray*}
There are similar expressions for $V_{q\bar{Q}}^{OGE}$ and
$V_{q\bar{Q}}^{conf}$. Because we preliminarily ignore the possible
hidden color contributions, we do not need $V_{Q\bar{Q}}^{OGE}$ and
$V_{q\bar{q}}^{OGE}$.

For the meson exchange potentials between two light quarks, we have
\cite{chiQM,ExchiQM}
\begin{eqnarray}
V_{uu}^{\sigma}(\bm{r}_{ij})&=&-C(g_{ch},m_\sigma,\Lambda)X_1(m_\sigma,\Lambda,r_{ij}),\\
V^{\pi_a}(\bm{r}_{ij})&=&C(g_{ch},m_{\pi_a},\Lambda)\frac{m_{\pi_a}^2}{12
m_{q_i}m_{q_j}}
X_2(m_{\pi_a},\Lambda,r_{ij})\nonumber\\
&&\times[\bm{\sigma}(i)\cdot\bm{\sigma}(j)][\tau_a(i)\tau_a(j)],\\
&&(a=1,2,3)\nonumber\\
V^{\rho_a}(\bm{r}_{ij})&=&C(g_{chv},m_{\rho_a},\Lambda)\left\{X_1(m_{\rho_a},\Lambda,r_{ij})
+\frac{m_{{\rho_a}}^2}{6
m_{q_i}m_{q_j}}\right.\nonumber\\
&&\times\left(1+\frac{f_{chv}}{g_{chv}}\frac{m_{q_i}+m_{q_j}}{M_N}+
(\frac{f_{chv}}{g_{chv}})^2\frac{m_{q_i}m_{q_j}}{M_N^2}\right)\nonumber\\
&&\times
X_2(m_{{\rho_a}},\Lambda,r_{ij})[\bm{\sigma}(i)\cdot\bm{\sigma}(j)]\Big\}[\tau_a(i)\tau_a(j)],\nonumber\\\\
V_{uu}^\omega(\bm{r}_{ij})
&=&C(g_{chv},m_\omega,\Lambda)\left\{X_1(m_\omega,\Lambda,r_{ij})
+\frac{m_{\omega}^2}{6
m_u^2}\right.\nonumber\\
&&\times\left(1+\frac{f_{chv}}{g_{chv}}\frac{2m_u}{M_N}+
(\frac{f_{chv}}{g_{chv}})^2\frac{m_u^2}{M_N^2}\right)\nonumber\\
&&\times
X_2(m_\omega,\Lambda,r_{ij})[\bm{\sigma}(i)\cdot\bm{\sigma}(j)]\left.\frac{}{}\right\}.
\end{eqnarray}
where
\begin{eqnarray}
C(g_{ch},m,\Lambda)&=&\frac{g_{ch}^2}{4\pi}\frac{\Lambda^2
m}{\Lambda^2-m^2},\\
X_1(m,\Lambda,r)&=&Y(mr)-\frac{\Lambda}{m}Y(\Lambda r),\\
X_2(m,\Lambda,r)&=&Y(mr)-\left(\frac{\Lambda}{m}\right)^3 Y(\Lambda r),\\
Y(x)&=&\frac{e^{-x}}{x}.
\end{eqnarray}

We do not present the tensor term and the spin-orbital term in the
meson exchange potentials since we consider only S-wave meson-meson
interactions. Here we use the same cutoff $\Lambda$ in describing
various meson interactions. Its value is around the scale of chiral
symmetry breaking ($\sim$1 GeV).

The interaction between a quark and an antiquark is related to that
between two quarks through the relation $V_{q\bar{q}}^M=G_M
V_{qq}^M$ where $G_M$ is the G-parity of the exchanged meson.

By calculating the RGM matrix elements and solving the RGM equation
for the bound state problem, one gets the energy of the system and
the relative motion wave function, from which one deduces the
binding energy $E_0=M_{\bar{Q}q}+M_{Q\bar{q}}-M_{sys}$. If $E_0$ is
positive, the system is bound.

For the model parameters, we take the values determined in the
previous investigations \cite{chiQM,ExchiQM}. The
harmonic-oscillator width parameter $b_u$=0.5 fm for $\chi$QM and
$b_u$=0.45 for E$\chi$QM. The up (down) quark mass $m_{u(d)}$=313
MeV. The coupling constant $g_{ch}$=2.621 is derived from the
measured $NN\pi$ coupling constant $g^2_{NN\pi}/4\pi=13.67$. The
masses of $\pi$, $\rho$, and $\omega$ are taken to be the
experimental values, whereas $\sigma$ meson mass is adjusted to fit
the binding energy of the deuteron. In E$\chi$QM, we use two sets of
values. We present the above parameters in Table \ref{paras}. The
parameters in the OGE and the confinement potentials can be derived
from the masses of the ground state baryons and the heavy mesons. In
fact, their values do not give effects to the binding energy of the
meson-antimeson system when we ignore the hidden color contributions
\cite{3872-rgm}. So we do not present their values here. For the
charm quark masses, we take $m_c$=1430 MeV \cite{zzz} and 1870 MeV
\cite{Semay} to see the heavy quark mass dependence of the binding
energy. For bottom quark, we use $m_b$=4720 MeV \cite{zzz2} and 5259
MeV \cite{Semay}. In our calculation, we take two values for the
cutoff $\Lambda$=1000 MeV and $\Lambda$=1500 MeV.

\begin{table}
\centering
\begin{tabular}{ccccc}
\hline
    & $\chi$QM & \multicolumn{2}{c}{E$\chi$QM } \\
    &        Set 1      & Set 2  & Set 3 \\\hline
$b_u$ (fm)& 0.5 & 0.45 & 0.45\\
$m_u$ (MeV)& 313 & 313 & 313\\
$m_\sigma$ (MeV)& 595 & 535 & 547\\
$g_{chv}$ &     & 2.351 & 1.972 \\
$f_{chv}/g_{chv}$ & & 0 & 2/3
\\\hline
\end{tabular}
\caption{Three sets of model parameters. Other meson masses are:
$m_\pi=138$ MeV, $m_\rho=775.8$ MeV, and $m_\omega=782.6$
MeV.}\label{paras}
\end{table}

\section{Results and discussions}\label{sec3}

\begin{figure}
\begin{center}
\begin{tabular}{c}
\includegraphics[angle=270,scale=0.32]{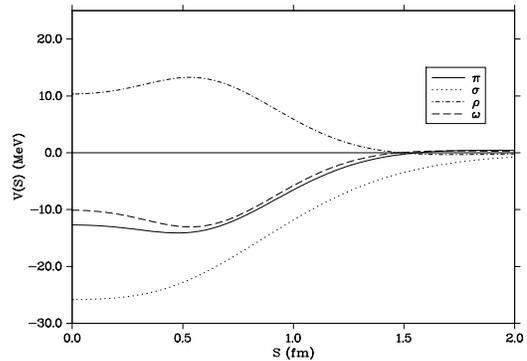}\\
(a)\\
\includegraphics[angle=270,scale=0.32]{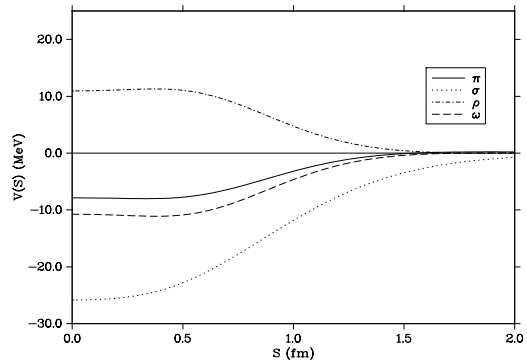}\\
(b)\\
\includegraphics[angle=270,scale=0.32]{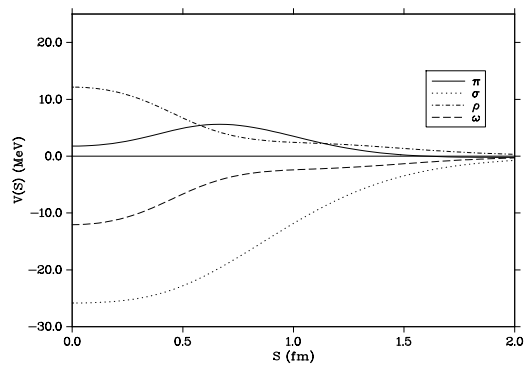}\\
(c)
\end{tabular} \caption{The meson exchange GCM matrix elements for
J=0 (a), J=1 (b), and J=2 (c) $D_1\bar{D}^*$ system. The used
parameters are $b_u$=0.45 fm, $m_\sigma$=547 MeV, $g_{chv}$=1.972,
$f_{chv}/g_{chv}$=2/3, $m_c$=1870 MeV, and $\Lambda$=1500
MeV.}\label{gcmME-D1}
\end{center}
\end{figure}

\begin{figure}
\begin{center}
\begin{tabular}{c}
\includegraphics[angle=270,scale=0.32]{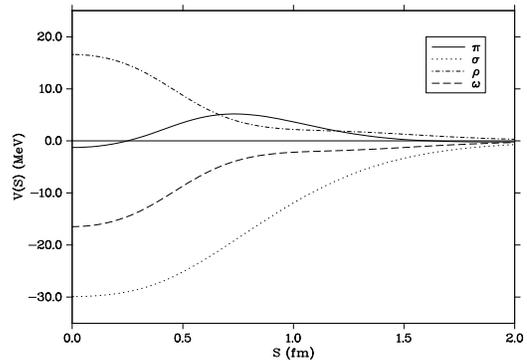}\\
(a)\\
\includegraphics[angle=270,scale=0.32]{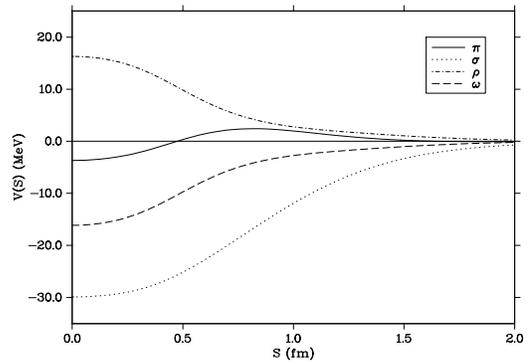}\\
(b)\\
\includegraphics[angle=270,scale=0.32]{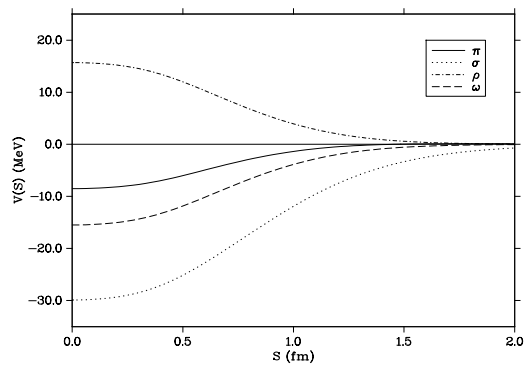}\\
(c)
\end{tabular}
\caption{The meson exchange GCM matrix elements for J=0 (a), J=1
(b), and J=2 (c) $D_1'\bar{D}^*$ system. The used parameters are the
same as those for $D_1\bar{D}^*$.}\label{gcmME-D1p}
\end{center}
\end{figure}

We first study the S-wave $D_1\bar{D}^*$ system. We illustrate the
diagonal meson exchange matrix elements of the Hamiltonian in the
generator coordinate method (GCM) calculation for different angular
momenta in Fig. \ref{gcmME-D1} with the parameters Set 3 in Table
\ref{paras}, $m_c$=1870 MeV and $\Lambda$=1500 MeV. One finds the
dominate contributions come from the $\sigma$ and $\pi$ meson
exchange interactions. The $J$=0 system is more attractive than
$J$=1, 2 systems. However, we do not find a binding solution in this
system with the parameters presented in the former section. For the
S-wave $D_1'\bar{D}^*$ system, the meson exchange GCM matrix
elements are illustrated in Fig. \ref{gcmME-D1p}. The $J$=2 system
is the most attractive one. But the system is also unbound. So our
preliminary calculation does not support the interpretation that
$Z^+(4430)$ is an S-wave $D_1\bar{D}^*$ or $D_1'\bar{D}^*$ bound
state.

\begin{figure}[htb]
\begin{center}
\includegraphics[angle=270,scale=0.32]{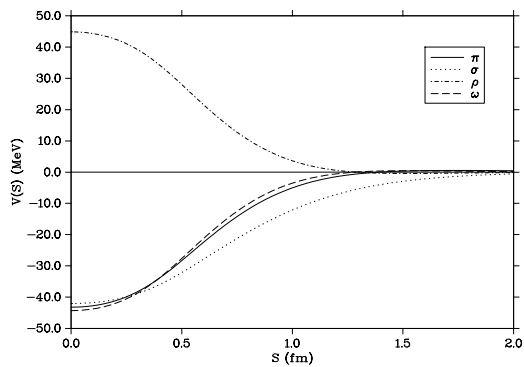}
\caption{The meson exchange GCM matrix elements for J=0
$D_1\bar{D}^*$ system with $G$=-. The used parameters are the same
as in Fig. \ref{gcmME-D1}.}\label{gcmME-D1-gm}
\end{center}
\end{figure}

The $D_1\bar{D}^*$ or $D_1'\bar{D}^*$ can also form a $G$=- system
with the flavor wave function
\begin{equation}
Z'=\frac{1}{\sqrt2}(\bar{D}_1^0D^{*+}-\bar{D}^{*0}D_1^+)
\end{equation}
or
\begin{equation}
Z'=\frac{1}{\sqrt2}(\bar{D}_1^{'0}D^{*+}-\bar{D}^{*0}D_1^{'+}).
\end{equation}
We found these systems are also unbound with our parameters. The
attractive force in this case is a little stronger than that in the
$G$=+ case. One observes this feature by comparing the GCM matrix
elements for $J$=0 case in Fig. \ref{gcmME-D1-gm} with those in
diagram (a) of Fig. \ref{gcmME-D1}.

The bottom analogs have better chances to form molecular states
because the kinetic term in the Hamiltonian has relatively small
contributions. We do get binding solutions with the parameters in
the former section. Table \ref{bottom} gives the binding energy and
the root-mean-square (RMS) radius for the $B_1\bar{B}^*$ and
$B_1'\bar{B}^*$ systems. The system is unbound for $J$=2
$B_1\bar{B}^*$ and $J$=0, 1 $B_1' \bar{B}^*$. In that table, we
present both the results for the $G$=+ case and those for the
$G$=$-$ case. A little deeper bound states appear in the later case.

\begin{table*}[htb]
\caption{The binding energy (RMS radius) for the bottom analog of
$D_1 \bar{D}^*$ and $D_1' \bar{D}^*$ system in unit of MeV (fm). A
$\times$ means the system is unbound. The system is unbound for
$J$=2 $B_1\bar{B}^*$ and $J$=0, 1 $B_1' \bar{B}^*$.}\label{bottom}
\begin{tabular}{ccc|ccc|ccc|ccc}\hline
           &        &          & \multicolumn{3}{c}{$B_1\bar{B}^*$ (J=0)}& \multicolumn{3}{c}{$B_1\bar{B}^*$ (J=1)}&\multicolumn{3}{c}{$B_1'\bar{B}^*$ (J=2)}\\
$G$-parity & $m_b$ (MeV)  & $\Lambda$ (MeV)& Set 1 & Set 2 & Set 3 & Set 1 & Set 2 & Set 3& Set 1 & Set 2 & Set 3  \\
$G$=+ &4720 & 1000 &1.0(1.4) &  3.7(1.2) &  3.4(1.2) &  $\times$ &  1.2(1.3) &  0.9(1.3) &   $\times$ &    0.6(1.3) &  0.3(1.3) \\
      &     & 1500 &1.6(1.4) &  4.8(1.2) &  4.4(1.2) &  $\times$ &  1.9(1.3) &  1.6(1.3) &   $\times$ &   1.3(1.3) &  1.0(1.3) \\
      &5259 & 1000 &1.8(1.3) &  4.8(1.1) &  4.5(1.1) &  $\times$ &  2.0(1.2) &  1.7(1.2) &   $\times$ &   1.4(1.2) &  1.1(1.2) \\
      &     & 1500 &2.4(1.3) &  5.9(1.1) &  5.5(1.1) &  0.4(1.4) &  2.9(1.2) &  2.5(1.2) &   $\times$ &   2.3(1.2) &  1.9(1.2) \\
$G$=- &4720 & 1000 &5.2(1.1) &  9.9(1.0) &  9.4(1.0) &  1.3(1.3) &  4.3(1.1) &  3.9(1.1) &   $\times$ &   1.3(1.3) &  0.9(1.3) \\
      &     & 1500 &6.9(1.1) & 13.0(0.9) & 12.4(0.9) &  2.2(1.3) &  6.0(1.1) &  5.5(1.1) &   $\times$ &   2.2(1.2) &  1.9(1.2) \\
      &5259 & 1000 &6.5(1.1) & 11.8(0.9) & 11.3(0.9) &  2.3(1.2) &  5.6(1.1) &  5.2(1.1) &   $\times$ &   2.2(1.2) &  1.9(1.2) \\
      &     & 1500 &8.4(1.0 & 15.1(0.9) & 14.5(0.9) &  3.3(1.2) &  7.5(1.0 &  7.0(1.0 &   0.6(1.3) &  3.3(1.1) &  2.9(1.1) \\
\hline
\end{tabular}
\end{table*}

Our study with the chiral quark model approach does not support the
existence of an S wave molecule in the $D_1\bar{D}^*$ and
$D_1'\bar{D}^*$ systems. This result is inconsistent with our
similar study at hadron level \cite{4430-xiang}. However, in the
case of $D\bar{D}^*$ system, we got consistent conclusions with
these two approaches \cite{3872-liu,3872-rgm}. A possible reason for
the present inconsistency is due to the different approximations in
getting the potentials. As a first step to derive the potential, one
writes out the quark-quark (or meson-meson) scattering matrix in
momentum space. The denominator of the propagator for a meson reads
$p^2-m^2+i\epsilon=p_0^2-\bm{p}^2-m^2+i\epsilon$ where $p$
($\bm{p}$) is the four(three)-momentum and $m$ is the meson mass.
The approximation $p^2\to -\bm{p}^2$, i.e. $p_0\sim0$, is adopted in
the chiral quark model approach, whereas the possible large $p_0$ is
considered for the hadron level calculation \cite{4430-xiang}. In
the later approach, the principal integration is always assumed if
$p_0$ is larger than the meson mass $m$ when we get the
coordinate-space potentials. In the present case, the large $p_0$ is
around $3m_\pi$ while the large $p_0$ is about $m_\pi$+7 MeV in the
$D\bar{D}^*$ case. Probably it is this $p_0$ around $3m_\pi$ leads
to inconsistent conclusions for the studies using these two
approaches. We reanalyzed the binding energies of the $D_1\bar{D}^*$
($D_1'\bar{D}^*$) system at hadron level with the approximation
$p_0\sim0$. As expected, we did not find a binding solution, which
indicates the important role of $p_0$. However, we need the
experiments to judge which approximation is correct. The comparison
of model predications with experimental measurements may finally
answer the puzzle.

Although our model calculation does not support the assumption that
$Z^+(4430)$ is a molecule, such an interpretation is still possible.
To get a more conclusive result in a future investigation, the
following effects may be included. First, the hidden-color
configuration and a larger model space may have contributions and
can be considered. Secondly, the coupling with D wave interaction is
probably not negligible and may be studied. Thirdly, the different
approximation in deriving the coordinate space potential may be
investigated in detail. In addition, our model neglects the
contribution from $\sigma$ exchange interaction between two heavy
quarks or between a heavy quark and a light quark. Although the
coupling constant $g_{QQ\sigma}$ is expected to be small, the value
may have big effects because no mass factor in the $\sigma$
potential can suppress the contribution. This is also an open
question one may discuss.

According to the GCM matrix elements, which roughly reflect the
force between the two mesons, if $Z^+(4430)$ can be identified as a
$D_1\bar{D}^*$ or $D_1'\bar{D}^*$ molecular state, a $G$=- state
around 4430 MeV should also exist. One expects that such a state may
be searched for in the $\pi^+\eta_c(2S)$, $J/\psi \pi^+\pi^0$, or
$\psi'\pi^+\pi^0$ channel.

In short summary, we have studied the bound state problem of the S
wave $D_1\bar{D}^*$ ($D_1'\bar{D}^*$) system in a chiral quark
model. Our preliminary calculation does not favor the assumption
that $Z^+(4430)$ is an S wave molecule. On the contrary, once
$Z^+(4430)$ ($G$=+) may be identified as a $D_1\bar{D}^*$
($D_1'\bar{D}^*$) molecule, its partner with $G=-$ should also
exist. When we move on to the bottom analogs, the existence of the
charged $B_1\bar{B}^*$ ($B_1'\bar{B}^*$) molecules with $G$=$+$ and
$G$=$-$ are both possible. Such states can probably be found in the
$\pi\Upsilon(2S)$, $\pi\eta_b(2S)$, $\Upsilon(1S) \pi^+\pi^0$, and
$\Upsilon(1S) \pi^+\pi^0$ channels in future measurements.

\section*{Acknowledgments}

YRL thanks Prof. S.L. Zhu, Prof. Q. Zhao, Prof. P.N. Shen for
helpful discussions and Prof. S. Olsen for correspondence. This
project was supported by the National Natural Science Foundation of
China under Grants 10775146 and 10805048, the Ministry of Science
and Technology of China (2009CB825200), the China Postdoctoral
Science foundation (20070420526), and K.C. Wong Education
Foundation, Hong Kong.

\end{document}